\documentclass[prb,superscriptaddress,showpacs,twocolumn]{revtex4}

\usepackage{graphicx}
\usepackage[latin1]{inputenc}
\usepackage{amsmath}

\newcommand{\eq}[1]{Eq.~(\ref{#1})}
\newcommand{\fig}[1]{Fig.~\ref{#1}}
\newcommand{\sect}[1]{Sec.~\ref{#1}}

\newcommand{\intg}[1]{\int\!\!\!d#1}
\newcommand{\sij}[1]{\sum_{\langle ij#1\rangle}}

\begin{document}

%%%%%%%%%%%%%%%%%%%%%%%%%%%%%%%%%%%%%%%%%%%%%%%%%%%%%%%%%%%%%%%%%%%%%%%%%%%%%%%
\title{Charge ordering in quarter-filled ladder systems coupled to the lattice}
%%%%%%%%%%%%%%%%%%%%%%%%%%%%%%%%%%%%%%%%%%%%%%%%%%%%%%%%%%%%%%%%%%%%%%%%%%%%%%%

\author{M. Aichhorn}
\author{M. Hohenadler}
\affiliation{Institut f\"ur Theoretische Physik,
Technische Universit\"at Graz,
Petersgasse 16, A-8010 Graz, Austria}

\author{E. Ya. Sherman}
\affiliation{Institut f\"ur Theoretische Physik,
Technische Universit\"at Graz,
Petersgasse 16, A-8010 Graz, Austria}
\affiliation{Institut f\"ur Theoretische Physik,
Karl-Franzens-Universit\"at Graz, Universit\"atsplatz 5, A-8010 Graz, Austria}

\author{J. Spitaler}
\author{C. Ambrosch-Draxl}
\affiliation{Institut f\"ur Theoretische Physik,
Karl-Franzens-Universit\"at Graz, Universit\"atsplatz 5, A-8010 Graz,
Austria}

\author{H. G. Evertz}
\affiliation{Institut f\"ur Theoretische Physik,
Technische Universit\"at Graz,
Petersgasse 16, A-8010 Graz, Austria}

\begin{abstract}
We investigate charge ordering in the presence of electron-phonon
coupling for  quarter-filled ladder systems by using exact diagonalization.
As an example we consider NaV$_2$O$_5$ using model parameters 
obtained from first-principles band-structure calculations.
The relevant Holstein coupling to the lattice considerably reduces the critical
value of the nearest-neighbor Coulomb repulsion at which formation of the
zig-zag charge-ordered state occurs, which is then accompanied by a static
lattice distortion. 
Energy and length of a kink-like 
excitation on the background of the distorted lattice are calculated. 
Spin and charge spectra on ladders with and without static distortion are obtained,
and the charge gap and the effective spin-spin exchange parameter $J$ are extracted.
$J$ agrees well with experimental results.
Analysis of the dynamical Holstein model, restricted to a small number of phonons,
shows that low frequency lattice vibrations have a strong influence on the
charge ordering, particularly in the vicinity of the phase transition
point. By investigating the charge order parameter we conclude that phonons
produce dynamical zig-zag lattice distortions. A model with only static
distortions gives a good description of the system well away from the
transition point while overestimating the amount of charge ordering in
the vicinity of the phase transition.
\end{abstract}

\pacs{71.10.Fd,71.38.-k,63.22.+m}

\maketitle

%%%%%%%%%%%%%%%%%%%%%%%%%%%%%%%%%%%%%%%%%%%%%%%%%%%%%%%%%%%%
\section{Introduction}
%%%%%%%%%%%%%%%%%%%%%%%%%%%%%%%%%%%%%%%%%%%%%%%%%%%%%%%%%%%%

The formation of an ordered pattern of ion charges is a rather 
general type of phase transition which occurs 
in three- as well as lower-dimensional solids. 
It has been  known for more than six decades
\cite{Vervey41} since its discovery in magnetite Fe$_3$O$_4$. Even in that
compound this phenomenon still attracts a lot of attention due to the interesting 
physics of the transition. Since the charge ordering causes changes in the interaction 
between the ions, it drives a lattice distortion, which, in turn,
influences the ordering pattern. The quarter-filled ladder compound NaV$_2$O$_5$ 
(Ref.~\onlinecite{Smolinski98})  
is another interesting example of a system which shows 
charge ordering. Here the transition, as observed in the
nuclear magnetic resonance experiments,\cite{Ohama99} occurs at $T_{\rm CO}\approx 35$ K
and is accompanied  
by the formation of a spin gap \cite{Isobe96} at the same 
\cite{Popova02} or slightly lower temperature.
In NaV$_2$O$_5$, where one $d_{xy}$ electron is shared by two V ions in a V-O-V rung,
the ordering occurs as a static charge disproportion $\delta$
between the V ions, which obtain charges $4.5\pm\delta$ 
with a zig-zag pattern of $\delta$'s.  Most probably, 
the main driving force for the transition is the Coulomb repulsion
of electrons on the nearest-neighbor sites within one leg of the ladder. 
For half the ladders the apical oxygen ion is located below the ladder
and for the other half above.
Therefore the crystal environment of the V ions is asymmetric, 
and the $d_{xy}$ electron is subject to a strong  Holstein-like electron-phonon coupling.
\cite{Sherman99}
As a result, the transition is accompanied by the displacement of ions from their 
positions in the high-temperature phase ($T>T_{\rm CO}$). They then form a larger
unit cell, as clearly seen directly in the x-ray diffraction experiments,\cite{Luedecke99}
where displacements of V ions of the order of 0.05 \AA\ were observed, 
and indirectly in the appearance of new phonon modes in the infrared absorption \cite{Popova97}
and Raman scattering spectra.\cite{Fischer99}

The importance of the coupling to the lattice for phase transitions in quarter-filled systems
was shown in Refs.~\onlinecite{Riera99,Clay03}.
The low-energy excitations of the zig-zag order parameter are 
also strongly influenced by the lattice.\cite{Sherman01}

The static properties of the ground state of the doped ladders without
coupling to lattice distortions were extensively 
investigated using mean field approaches,\cite{Seo98,Thalmeier98,Mostovoy00} 
the density-matrix renormalization-group (DMRG),\cite{Vojta01}
bosonization, and renormalization group techniques.\cite{Orignac03}
The DMRG studies show that at strong 
enough Hubbard interaction the ladders exhibit a zig-zag charge order
if the repulsion between the electrons on the nearest-neighbor sites exceeds
some critical value $V_c$. The corresponding phase transition 
is of second-order as a function of $V$.
For a very strong intersite repulsion, phase separation becomes possible.\cite{Vojta01}
Dynamical properties were studied with exact numerical diagonalization
\cite{Aichhorn02,Hubsch01,Cuoco99} without taking into account the coupling to the lattice. 
They were quite successful in understanding NaV$_2$O$_5$ dynamical properties 
above the transition temperature.

The investigation of the ordering of electrons interacting with 
the lattice requires the knowledge
of electron-phonon couplings and lattice force constants in addition
to electronic parameters such as hopping matrix 
elements and electron correlations.  For a given compound  
almost all of these parameters can be extracted from 
first-principles band-structure calculations. 
The force constants and electron-phonon coupling 
can be obtained by comparing  the total energies
and the interionic forces in distorted and undistorted lattices.
The phonon frequencies required for studies of dynamical
lattice distortions  are given by experiment,\cite{Fischer99} while
the necessary knowledge of their eigenvectors can be obtained 
from first-principles calculations.   
We performed such calculations of
the band structure, lattice dynamics and electron-phonon 
coupling for the NaV$_2$O$_5$ compound. 
These calculations are in good agreement with 
Ref.~\onlinecite{Smolinski98} in the part concerning the band structure,
and their details as well as a comparison to other 
first-principles calculations \cite{fpc} will be published elsewhere.\cite{Spitaler03}
In the present paper we will concentrate on the strongest 
electron-phonon mode present in NaV$_2$O$_5$, which is a simple Holstein-type interaction.\cite{Spitaler03} 

We therefore investigate a model which takes into account the main interactions, 
i.e., the Hubbard and intersite repulsions and the coupling to the lattice. 
We study the ground-state properties of a quarter-filled ladder 
coupled to static lattice distortion with the Lanczos algorithm.
We show that the ordering is strongly enhanced by the interaction with the lattice for realistic 
values of the coupling. Secondly, the interaction with dynamical (quantum)
phonons at different frequencies within the Holstein model is considered.
We find that indeed, the quantum phonons produce and stabilize the static order.

The paper is organized as follows. In \sect{sec:model} we introduce the
extended Hubbard model (EHM) with additional lattice distortions. In
\sect{sec:statics} we present results for static properties of this model,
including kink excitations. Dynamical quantities are discussed in
\sect{sec:dynamics}. Section \ref{sec:holstein} focuses on the influence of
dynamical phonons, and finally we give our conclusions in \sect{sec:conclusions}.

%%%%%%%%%%%%%%%%%%%%%%%%%%%%%%%%%%%%%%%%%%%%%%%%%%%%%%%%%%%%
\section{Model}\label{sec:model}
%%%%%%%%%%%%%%%%%%%%%%%%%%%%%%%%%%%%%%%%%%%%%%%%%%%%%%%%%%%%

\begin{figure}
  \centering
  \includegraphics[width=0.35\textwidth]{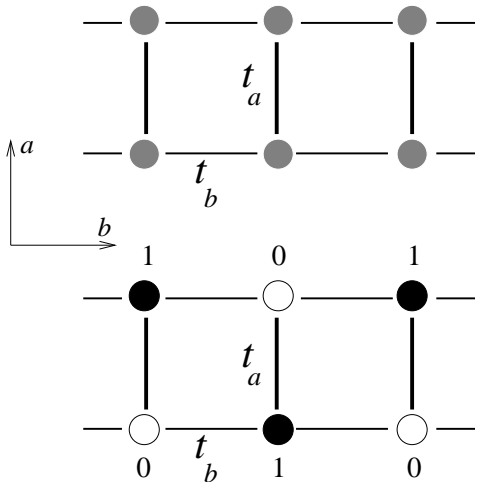}
  \caption{\label{fig:ladders}%
           Schematic picture of ladders, with hopping matrix
           elements $t_a$ on the rungs and $t_b$ along the chains.
           The darkness of the circles corresponds to the charges on the
           sites of the ladder. 
           The upper ladder is shown without charge order, 
           the lower ladder with zigzag charge order.
                }
\end{figure}

The quarter-filled ladder compound $\alpha^\prime$-NaV$_2$O$_5$ can be  described 
microscopically by an extended Hubbard (or $t$-$U$-$V$)  model
(EHM). For the description of the distorted low-temperature phase
we also include  the coupling of electrons to the lattice, 
yielding the model
\begin{equation}\label{ham}
    H=H_{\rm EHM}+H_{ l}+H_{ e-l},
\end{equation}
where $H_{ l}$ is the 
lattice deformation contribution, and $H_{ e-l}$ the electron-lattice
interaction. These terms are given by
\begin{subequations}
\begin{align}
    H_{\rm EHM}=&-\sum_{\langle ij\rangle,\sigma}t_{ij}\left(c_{i\sigma}^\dagger
    c_{j\sigma}^{\phantom{\dagger}}+\mbox{H.c.}\right)\label{hehm}\nonumber\\
    &+U\sum_in_{i\uparrow}n_{i\downarrow}+\sij{}V_{ij}n_in_j,\\
    H_{l}= &\kappa\sum_i\frac{z_i^2}{2},\label{eq_2b}\\
    H_{e-l}=&-C\sum_iz_in_i,\label{eq_2c}
\end{align}
\end{subequations}
with the effective lattice force constant $\kappa$ and the Holstein constant
$C$. The sites are labeled by the indices $i$, $j$ and $z_i$ is the
distortion on site $i$. The hopping matrix elements $t_{ij}$ connect nearest
neighbor sites $\langle ij\rangle$ (see \fig{fig:ladders}) 
with occupation numbers $n_i = n_{i\uparrow} + n_{i\downarrow}$.

The first-principles calculations done in
Ref.~\onlinecite{Smolinski98} and by our group\cite{Spitaler03}   
yield the intrarung hopping $t_a\approx0.35$ eV, which we will 
use below as the unit of energy.  For the hopping along the ladder we use 
$t_b=t_a/2$, again in agreement with the band-structure 
results,\cite{Smolinski98,Spitaler03}  
whereas previous DMRG studies\cite{Vojta01} were mostly done at $t_a/t_b\le 1.4$. 
For the on-site  Coulomb interaction we use $U=8.0$ 
as estimated in Ref.~\onlinecite{Smolinski98}.
We assume $V_{ij}=V$ to be the same for all bonds, and take $V$ as a free parameter
since there is no unique procedure of extracting it from the 
band-structure calculations. 
The lattice distortions are expressed in units of 0.05\,\AA\  since the
ion displacement below $T_{\rm CO}$ are of this order of magnitude.\cite{Luedecke99}  
With the chosen units of energy and length the comparison
of the band structure and lattice force calculations 
done on distorted and undistorted lattices give 
the dimensionless constants $\kappa=0.125$ and $C=0.35$,
respectively.\cite{Spitaler03} The effective coupling parameter
$C^2/\kappa$ is close to unity and, therefore, the lattice plays an important role
in determining the properties of NaV$_2$O$_5$. 

The Hamiltonian in \eq{ham} will be used for calculations with both
static and dynamical lattice distortions. All quantities presented in this paper 
are calculated by the ground state Lanczos method 
on single ladders of up to eight rungs, 
which enabled us to perform simple finite size scaling. The largest Hilbert
space for the eight rung lattice considered in this study was of dimension
$N_{\rm states}=1\,656\,592$, which could be reduced in special cases by
exploiting translational invariance and $S_z$ conservation to $N_{\rm
  states}=103\,820$. The next lattice size admitting charge order would
consist of ten rungs, which is far beyond our computational capabilities.
The restriction to a single ladder is a considerable
simplification compared to the structure of NaV$_2$O$_5$, 
but for the quantities of interest in this paper the
role of other ladders is of minor importance, since the frustration of
the ladder-ladder interactions significantly reduces their effect
in the zig-zag ordered state.

%%%%%%%%%%%%%%%%%%%%%%%%%%%%%%%%%%%%%%%%%%%%%%%%%%%%%%%%%%%%
\section{Static properties}\label{sec:statics}
%%%%%%%%%%%%%%%%%%%%%%%%%%%%%%%%%%%%%%%%%%%%%%%%%%%%%%%%%%%%

%%%%%%%%%%%%%%%%%%%%%%%%%%%%%%%%%%%%%%%%%%%%%%%%%%%%%%%%%%%%
\subsection{Charge order}\label{sec:order}
%%%%%%%%%%%%%%%%%%%%%%%%%%%%%%%%%%%%%%%%%%%%%%%%%%%%%%%%%%%%
To investigate the connection between the lattice distortion
and charge ordering we calculated the static charge structure factor
\begin{equation}
    S_c({\mathbf q})=\frac{1}{N}\sum_{ij}e^{i{\mathbf
    q}({\mathbf R}_i-{\mathbf R}_j)}\left(\langle n_i n_j\rangle - 
       \langle n\rangle^2 \right),
\end{equation}
where $N$ is the total number of sites in the system. 
The zigzag charge order parameter $m_{\rm CO}$ can be expressed in terms of
this structure factor as 
\begin{equation}\label{msqr}
    m_{\rm CO}^2=\frac{1}{N\langle n\rangle^2}S_c({\mathbf Q}),\quad {\mathbf Q}=(\pi,\pi).
\end{equation}
The term $\langle n\rangle^2$ in the denominator ensures that the
order parameter is equal to unity for full ordering, which in
NaV$_2$O$_5$ corresponds to the charges of V ions within a rung  
to be equal to $+5$ and $+4$, respectively.

\begin{figure}[t]
  \centering
  \includegraphics[width=0.45\textwidth]{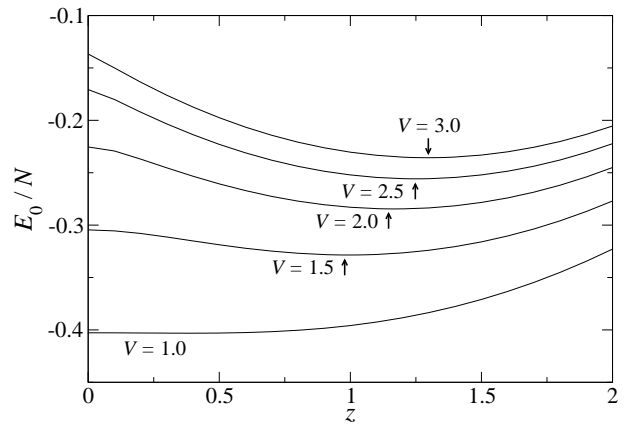}
  \caption{\label{fig:E0}%
    Ground-state energy per site as a function of the distortion $z$ calculated
    on an $8 \times 2$ system with periodic boundary conditions along the ladder,
    and $C=0.35$.
    From bottom to top: $V=1.0$, $1.5$, 
    $2.0$, $2.5$, $3.0$. The arrows indicate the position of the minimum.}
\end{figure}

At this point the lattice distortions $z_i$ in the Hamiltonian are 
external parameters of the model and not dynamical variables.
Therefore they have to be fixed in a proper way, for which
we chose a mean field approach. Considering the distortions as
the mean field parameters one can extract the optimal value for the $z_i$
by looking for the minimum in the ground-state energy with respect
to $z_i$. This procedure could be done within the unrestricted
Hartree-Fock approximation, but this complicates the calculation
because of the larger number of variables for which the minimum
has to be found. Instead we restrict ourselves to a single 
order pattern, the zig-zag order,\cite{Seo98,Mostovoy00}  
which was observed experimentally:\cite{Luedecke99}
\begin{equation}\label{eq_5}
    z_i=ze^{i{\mathbf Q}\cdot{\mathbf R}_i}.
\end{equation}
and investigate the total energy as a function of $z$. The optimal values of $z$ 
where the total energy reaches the minimum for several values of the
nearest-neighbor Coulomb interaction $V$ determined in this way  
are indicated by arrows in \fig{fig:E0}. In the following we denote the
position of the minimum in the ground-state energy by $z_{\rm min}$.

\begin{figure}[t]
  \centering
  \includegraphics[width=0.45\textwidth]{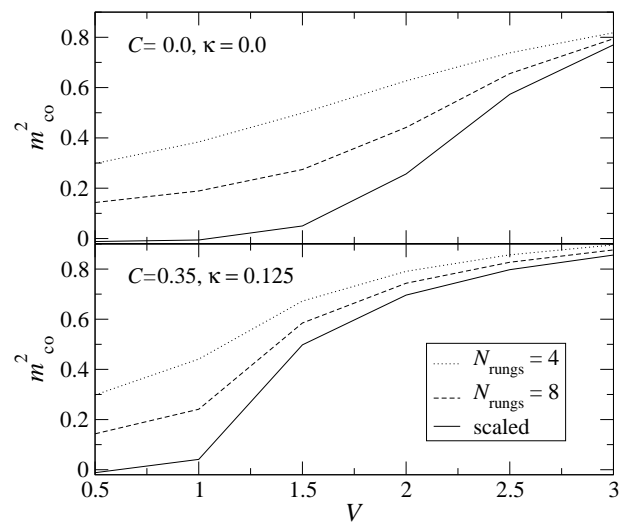}
  \caption{\label{fig:orderpar}%
    Charge order parameter for several values of $V$. Upper/lower panel:
    Calculation without/with coupling to the lattice.
    The solid lines are obtained by $1/N_{\rm rungs}$ finite size extrapolation.
   }
\end{figure}
\begin{figure}[t]
  \centering
  \includegraphics[width=0.45\textwidth]{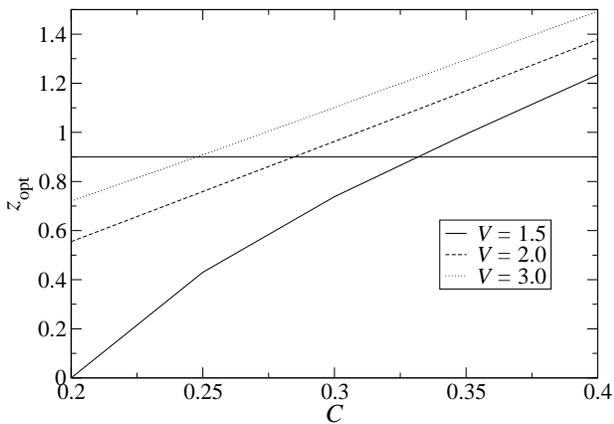}
  \caption{\label{fig:optdist}%
    Optimal distortion $z_{\rm opt}$ as a function of the Holstein constant $C$ for
   different values of the coulomb interaction $V$, calculated on a $8\times
   2$ cluster. The horizontal line indicates the experimental result.
   }
\end{figure}
\begin{figure}[t]
  \centering
  \includegraphics[width=0.45\textwidth]{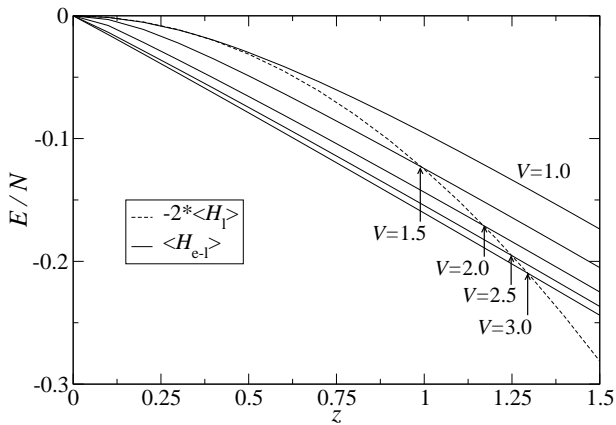}
  \caption{\label{fig:virialth}%
    Contribution of the lattice energy \eq{eq_2b} (dashed) and the electron-lattice
    energy \eq{eq_2c} (solid) to the ground state energy as a function of the
    distortion $z$, at $C=0.35$. The lattice energy is independent of $V$. 
    Electron-lattice energy, from top to bottom: $V=1.0$, $1.5$, $2.0$, $2.5$, $3.0$.
    The arrows are drawn where the { total} energy has its minimum, as in \fig{fig:E0}.
    }
\end{figure}

For interaction strengths of $V=1.5$ up to $V=3.0$ a clear minimum occurs at 
$z\approx 1$. Additionally one finds a maximum at $z=0$, which results from
the $z\to -z$ symmetry  of the system. We also found a small 
distortion for $V=1.0$, but it is strongly
size dependent and rapidly decreases with increasing length of the ladder,
whereas the distortions marked by arrows in \fig{fig:E0} are almost
independent of the system size. Therefore we argue that the finite value of $z$ at $V=1.0$
is due to finite-size effects and disappears in the thermodynamic
limit. For this reason we did not mark it with an arrow in \fig{fig:E0}.

This behavior gives a first idea about the charge ordering of the
system with and without coupling to the lattice. From \fig{fig:E0} one could expect 
that the charge ordering transition
in the presence of static mean-field-like lattice distortions
occurs in the region between $V=1.0$ and $V=1.5$. In order to investigate
this transition we calculate the order parameter given by \eq{msqr}. This
quantity shows strong finite-size effects, which makes it necessary to apply
finite-size scaling. We calculated the order parameter for systems of four
and eight rungs, respectively, the largest system size available.
Although higher-order corrections to the scaling
behavior are expected for these small systems, we performed a $1/N_{\rm
  rungs}$ extrapolation to $1/N_{\rm rungs}=0$. This
procedure does not give the exact value of the order parameter in the
thermodynamic limit and does not allow to extract an exact value for the
critical Coulomb interaction $V_c$, but it provides the possibility to obtain
a rough estimate of the interaction $V$ at which the phase transition
occurs as well as the approximate $m_{\rm CO}^2(V)$-dependence.

To study the effects of the lattice distortions we also 
calculated the order parameter without coupling to the lattice, that is
for $C=0$. The results are shown in \fig{fig:orderpar}. 
As one can see in the upper
panel where no lattice distortions are present, 
the order parameter changes rather
smoothly when going from the disordered phase into the ordered one. A
different behavior can be found for finite lattice distortions. Here the
charge ordering sets in, at a much lower value of $V$
than in the absence of distortions. In addition the transition
sharpens considerably. For both zero and finite distortions the finite size scaled
order parameter is slightly negative for small interactions $V$, which is due to the
fact that higher order corrections in the scaling have been neglected.

The optimal distortions $z$ for
different values of the Holstein constant $C$ and repulsion  $V$ are presented 
in \fig{fig:optdist}. In our units the distortion
found experimentally in the
charge-ordered phase is at
approximately $z=0.9$,\cite{Luedecke99} indicated by a horizontal line in \fig{fig:optdist}.
For $V=1.5$, close to the charge order phase transition,
this value of $z=0.9$ is reached near $C=0.33$. 
The distortion close to $z=0.9$ is realized also for other interactions,
e.g. deep in the ordered phase at  $V=3.0$, $C=0.24$.
Since, however, NaV$_2$O$_5$ is probably close to a quantum
critical point of charge ordering,\cite{Cuoco99}
and because of the value for $C$ obtained
from the band structure calculations, 
we conclude that the system can best be
described in the ordered phase using $V\approx 1.3$ and $C\approx 0.35$. 

When calculating the ground-state energy of the system, the question
arises how the terms in \eq{ham} contribute to the total energy,
i.e., whether some sort of virial theorem holds.
In \fig{fig:virialth} the behavior of the two
contributions (\ref{eq_2b}) and (\ref{eq_2c}) is
shown. One can easily see that the crossing points of the curves are at the
same value of $z$ at which the ground-state energy reaches its minimum,
\fig{fig:E0}. 
Therefore we find that a virial theorem such as that for the
one-electron polaronic states,\cite{Rashba82} is fulfilled in the form 
\begin{equation}
 \langle H_{l}\rangle=-\frac{1}{2}\langle H_{ e-l}\rangle,
\end{equation}
with a relative numerical accuracy of better than $10^{-4}$.
For large Coulomb interactions well above the phase transition
this high accuracy is likely achieved since the ion charges depend very
weakly on $z$. Therefore, compared to $H_{e-l}$ and $H_{l}$, $H_{\rm
  EHM}$ is almost independent of $z$, namely $\left[dH_{\rm EHM}/dz(z_{\rm
    min})\approx 0.005\right]$, and the dependence of 
  $H_{e-l}(z)$ in \fig{fig:virialth} is close to a straight line. Then the
  virial 
theorem follows from the functional form of $H_{l}(z)$ and $H_{e-l}(z)$. For
smaller values of the interaction, e.g., $V=1.5$, the 
dependence of the ion charges and $H_{\rm EHM}$ is considerably larger
$\left[dH_{\rm EHM}/dz(z_{\rm min})\approx
  0.03\right]$. Yet also in this case the virial relation is satisfied.  
The contribution of the sum of the lattice terms 
$\langle H_{l}+H_{e-l}\rangle$ to the total energy varies
between 19\% at $V=1.5$ and 30\% at $V=3.0$.

%%%%%%%%%%%%%%%%%%%%%%%%%%%%%%%%%%%%%%%%%%%%%%%%%%%%%%%%%%%%
\subsection{Kink excitations}\label{sec:kinkex}
%%%%%%%%%%%%%%%%%%%%%%%%%%%%%%%%%%%%%%%%%%%%%%%%%%%%%%%%%%%%

\begin{figure}[t]
  \centering
  \includegraphics[width=0.45\textwidth]{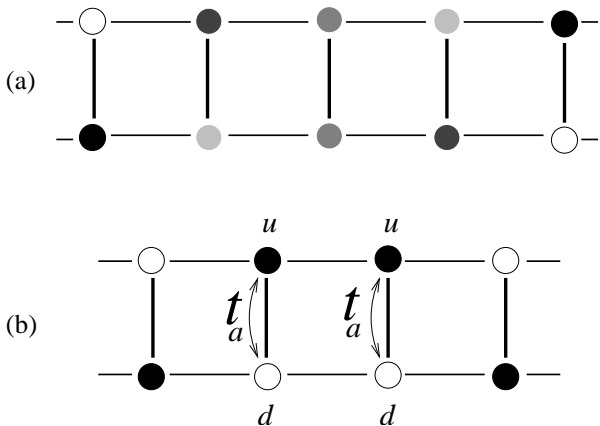}
  \caption{\label{fig:kinks}%
   (a) A schematic plot of a non-local kink-like excitation in an ordered
   ladder. 
   The darkness of the circles corresponds to the charges on the
   sites of the ladder. 
   (b) Local kink excitation with a sharp change of the
   order  parameter. The electron states for one of the electrons moving between
   sites $u$ and $d$  are degenerate in this case.} 
\end{figure}
\begin{figure}[t]
  \centering
  \includegraphics[width=0.45\textwidth]{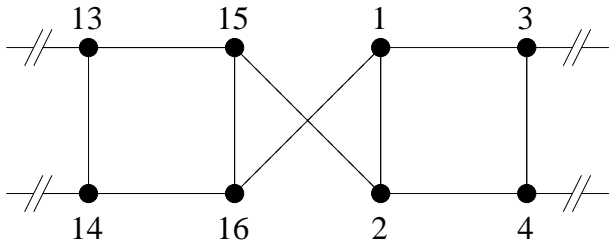}
  \caption{\label{fig:twistedbc}%
    Twisted ("M\"{o}bius") boundary conditions which produce a kink excitation. The
  numbers label the sites in the $8 \times 2$ cluster from 1 to 16.
  }
\end{figure}

So far we have considered the perfect zig-zag charge order pattern
described by \eq{eq_5}. This ordering can be destroyed by local in-rung excitations 
where an electron hops from the site with minimal energy to that with maximal energy,
as marked by black and white circles in \fig{fig:ladders}.
The excitation energy of this
process is $2V$, provided the system is totally ordered, that is, 
$m_{\rm CO}=1.0$. Another type of excitation is the formation of a local pair of
doubly occupied and empty rungs, which has the same energy $2V$ at $m_{\rm CO}=1.0$. 
In addition, there are nonlocal kinklike excitations, where the
order parameter smoothly changes along the ladder between two degenerate
patterns as shown in \fig{fig:kinks}.  
The nonlocal character of the
excitation leads to a decrease of the excitation energy. Since the lattice is coupled
to ion charges by the Holstein interaction, the kinks couple to the lattice, too.
 
\begin{figure}[t]
  \centering
  \includegraphics[width=0.45\textwidth]{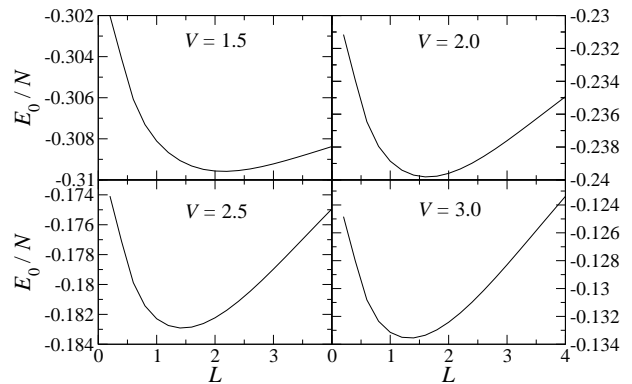}
  \caption{\label{fig:kinkenerg}%
    Ground-state energy per site as a function of the kink length $L$, \eq{eq_7}.} 
\end{figure}

In order to investigate kink excitations we used the largest system size
available, which is a single ladder consisting of eight rungs, and imposed
twisted "M\"{o}bius" boundary conditions \cite{Okunishi01}
as shown in \fig{fig:twistedbc}. The zig-zag distortions of
\eq{eq_5} are modified to a kink distortion
\begin{equation}\label{eq_7}
  z_i=ze^{i{\mathbf{Q\cdot R}}_i}\tanh\left[\frac{({\mathbf R}_i-{\mathbf
  R}_0)\hat{\mathbf e}_a}{L}\right],
\end{equation}
with the center of the kink ${\mathbf R}_0$  located in the middle between the
rungs, $L$ being its length in units of
the lattice spacing, and  $\hat{\mathbf e}_a$ the unit vector in ladder
direction. 
We kept $z$ at its previous optimal value $z_{\rm min}$ and varied $L$
looking for the minimum in the ground-state energy as shown in
\fig{fig:kinkenerg}.

\begin{figure}[t]
  \centering
  \includegraphics[width=0.45\textwidth]{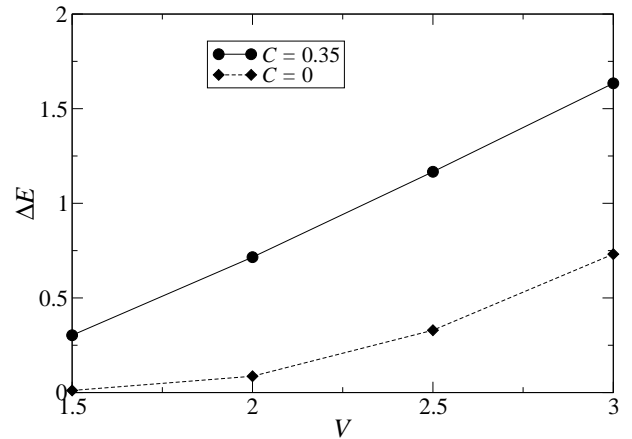}
  \caption{\label{fig:kink_exitenerg}%
    Kink excitation energy as a function of the interaction $V$.
    It is zero for $V\le 1.0$. 
    The lines are guides to the eye.
   }
\end{figure}

For interactions up to $V=1.0$ the ground-state energy strongly decreases
with increasing kink length $L$, without a minimum,
implying that at this weak coupling we have no kink excitations in the system. 
For larger interactions we
found a clear minimum in the ground-state energy. 
The kink length at $V=1.5$ is $L\approx 2.15$,
and it shrinks to $L\approx 1.33$ at $V=3.0$. 

The kink excitation energy is defined as the difference between the
ground-state energy with twisted boundary conditions at the optimal value of
$L$ and 
the ground-state energy with periodic boundary conditions and static
distortions $z=z_{\rm min}$. It is shown in \fig{fig:kink_exitenerg}. 
In the fully ordered state and in the atomic limit where $V\gg t_a$,
the kink potential energy $\Delta E$ is close to $V$ (see \fig{fig:kinks}). 
The actual total energy is considerably
smaller since the kinks are extended and since 
at the kink boundary the electron energy on the upper ($u$)
and lower ($d$) legs become degenerate as shown in \fig{fig:kinks}, 
and therefore the intrarung hopping 
becomes more likely. The hopping kinetic energy of the order of
$-t_a$ decreases the total energy of the system, leading to the result
shown in \fig{fig:kink_exitenerg}.

For weak interactions, where the kinks are well 
extended and $m_{\rm co}^2\ll 1$, 
our results can be compared with a model calculation in a 
classical $\phi^4$ model for infinite ladders\cite{Sherman01} which gives
\begin{subequations}
\begin{align}
L&=\frac{1}{\sqrt{V-V_c}}\label{eq:klength},\\
\Delta E&=\frac{3}{2}V\left(V-V_c\right)^{3/2},\label{eq:kenergs}
\end{align}
\end{subequations}
where $V_c$ is the critical value of the Coulomb interaction for the phase
transition. To make a connection to our results,
we estimate $V_c$ from Eqs.~(\ref{eq:klength})
and (\ref{eq:kenergs}) for the distorted lattice at $V=1.5$ and $C=0.35$
independently and compare them. From $L$ and  
$\Delta E$ we obtain $V_c=1.28$ and $V_c=1.05$, respectively. These
values are in reasonable agreement with each other and consistent  
with the behavior of the order parameter, see \fig{fig:orderpar}.

At all values of $V$ the kink excitation energy with lattice coupling is
larger than the excitation energy without coupling shown as dashed
line in \fig{fig:kink_exitenerg}. This can be understood since at a fixed
value of $V>V_c$ the charge order parameter is larger for the distorted
lattice. Since the ordering is more complete, the kink lengths are smaller which
increases the excitation energy. Note that without lattice coupling we have
no parameter $L$ as in \eq{eq_7} for the determination of the kink
length.

%%%%%%%%%%%%%%%%%%%%%%%%%%%%%%%%%%%%%%%%%%%%%%%%%%%%%%%%%%%%
\section{Dynamic properties}\label{sec:dynamics}
%%%%%%%%%%%%%%%%%%%%%%%%%%%%%%%%%%%%%%%%%%%%%%%%%%%%%%%%%%%%

\begin{figure}[t]
  \centering
  \includegraphics[width=0.45\textwidth]{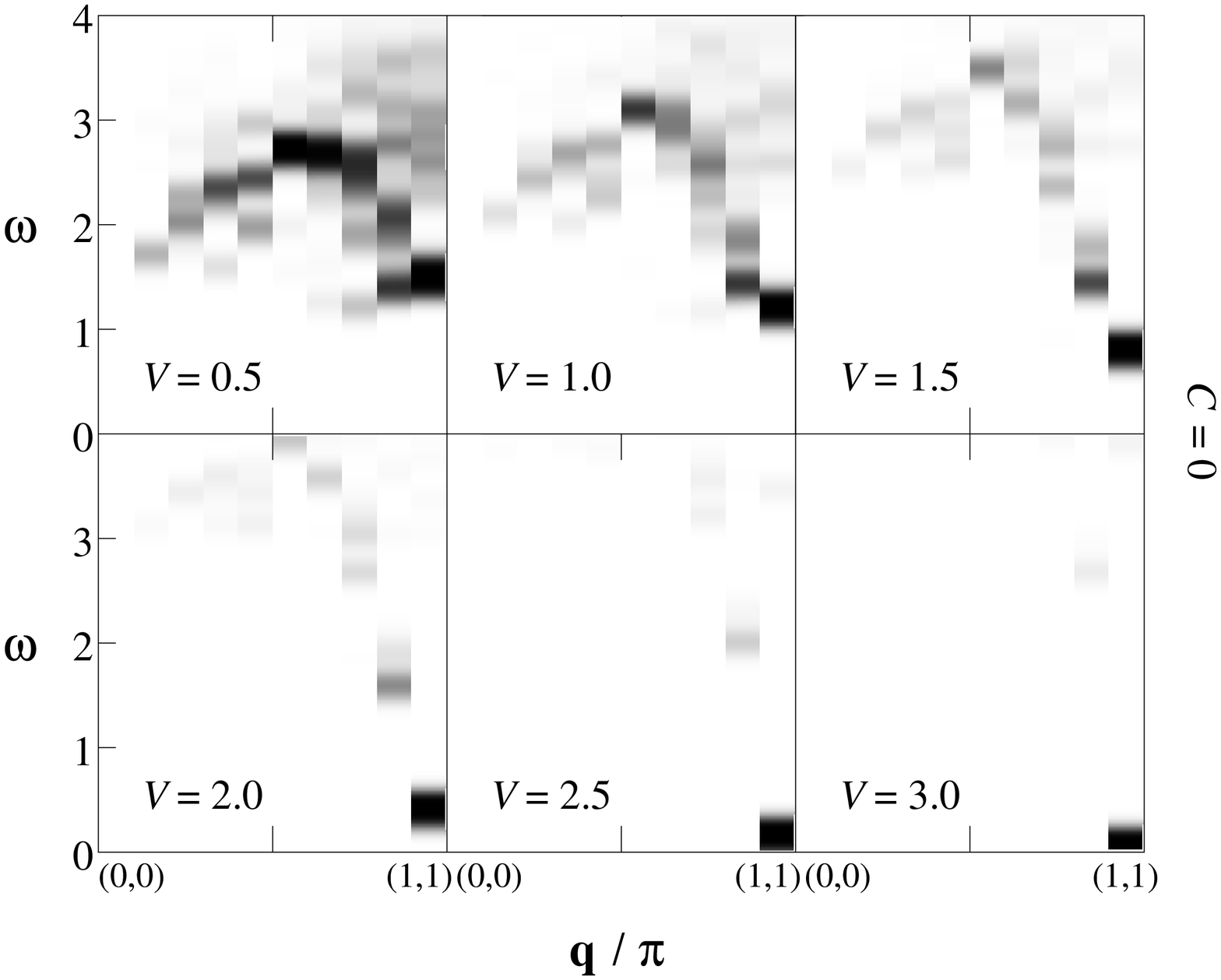}\\
  \includegraphics[width=0.45\textwidth]{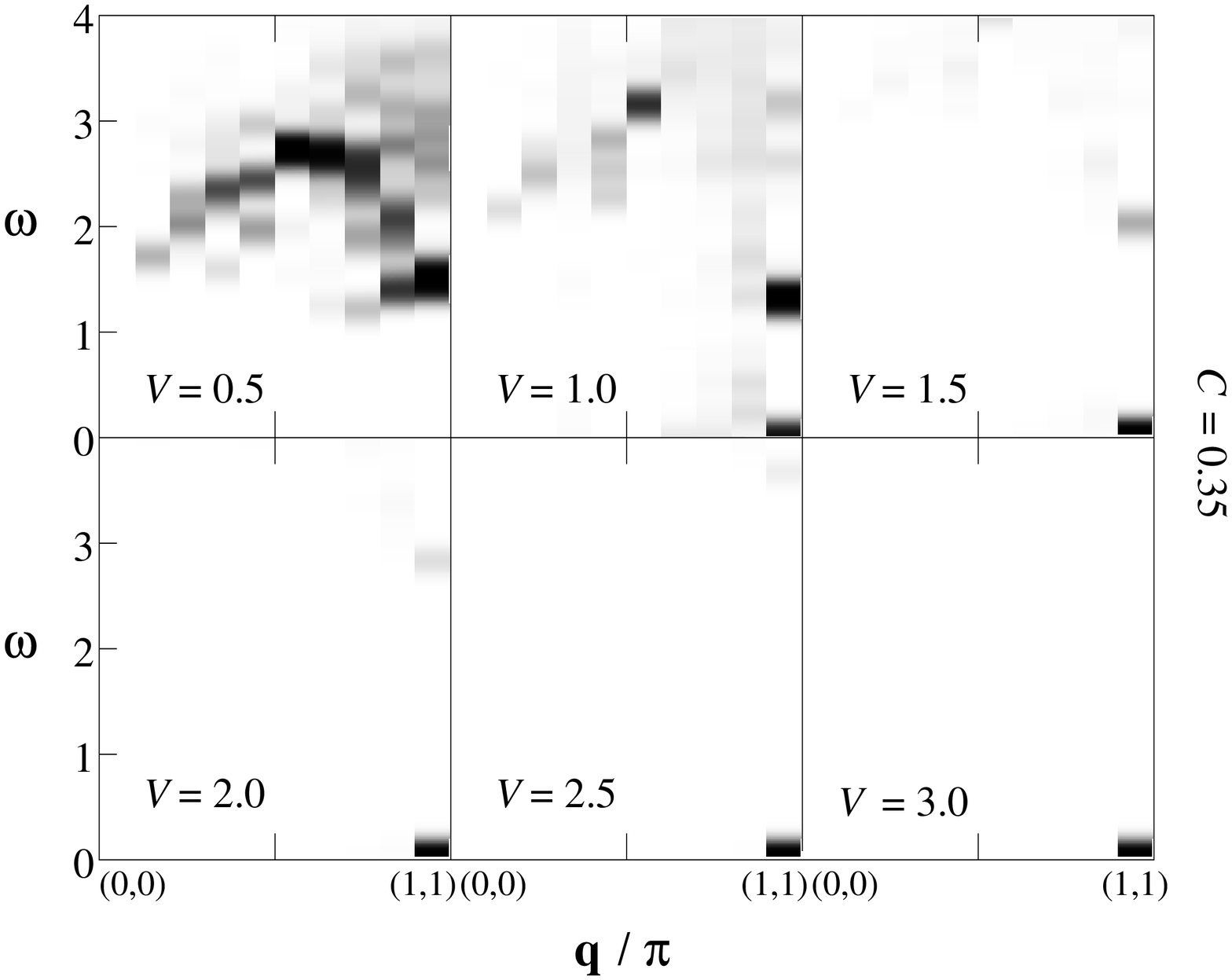}
  \caption{\label{fig:chargesusc}%
    Charge susceptibility calculated on an $8\times 2$ ladder. Upper panel: without lattice
    coupling. Lower panel: With lattice coupling ($C=0.35$). The wave vector scan consists
    of the two ranges $(q_a,q_b)/\pi=(0,0)\to(0,1)$ and $(1,0)\to(1,1)$, 
    separated by a tick mark on the horizontal axis.
    An additional broadening of width $\eta=0.1$ was used.}
\end{figure}
\begin{figure}[t]
  \centering
  \includegraphics[width=0.45\textwidth]{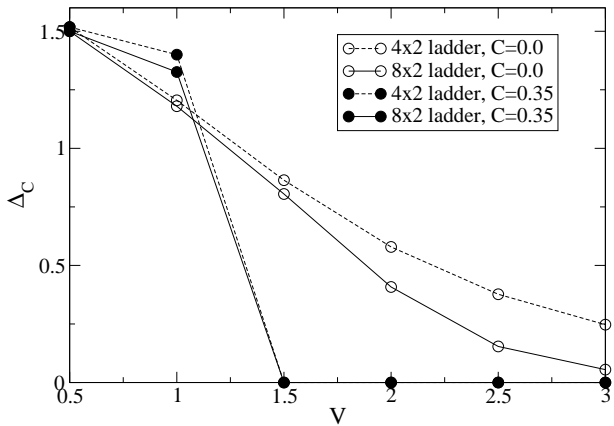}
  \caption{\label{fig:chgap}%
    Charge gap $\Delta_C$ at ${\mathbf q}={\bf Q}$ of the charge
    susceptibility as a function of $V$. Solid lines: $8\times 2$
    ladder. Dashed lines: $4\times 2$ ladder. Open symbols: Without lattice
    coupling. Full symbols: With lattice coupling (see text for case $V=1.0$).
  }
\end{figure}

In \sect{sec:statics} we only  considered static properties.
However, enlightening insight into
the physics of the system can be extracted from dynamical correlation
functions showing the spectra of 
charge and spin excitations. The corresponding susceptibilities are given by
\begin{subequations}
\begin{align}
  \chi_{C}({\mathbf q},\omega)&=\intg{t}e^{i\omega t}\langle n_{\mathbf
    q}(t)n_{-{\mathbf q}}\rangle\label{eq:chsusz}\\
  \chi_{S}({\mathbf q},\omega)&=\intg{t}e^{i\omega t}\langle S^z_{\mathbf
    q}(t)S^z_{-{\mathbf q}}\rangle,
   \label{eq:spinsusz}
\end{align}
\end{subequations}
where $n_{\mathbf q}(t), n_{-{\mathbf q}}$ and $S^z_{\mathbf q}(t),S^z_{-{\mathbf q}}$  are the 
the Fourier transforms of the charge and $z$ component 
of spin densities, respectively.

We calculated the charge susceptibility \eq{eq:chsusz} on a ladder consisting of
eight rungs with periodic boundary conditions along the ladder. The results are
shown in \fig{fig:chargesusc}. We define the charge gap $\Delta_C$ as
the energy at which the lowest lying excitation of the charge susceptibility
occurs. 
The corresponding momentum is always ${\mathbf q}={\bf Q}$. 
In the disordered phase we have no gapless charge excitation. 
When increasing the Coulomb interaction $V$, the gap at
${\mathbf q}={\bf Q}$ decreases as shown in \fig{fig:chgap},
and all other charge excitations become insignificant.
The charge gap
does not vanish exactly for $C=0$, i.e. without coupling to the lattice, 
but appears to go to zero as $N_{\rm rungs}\to\infty$.
When the coupling to the lattice is switched on,
the gap is exactly zero in the ordered phase where the symmetry is
broken explicitly. 
The charge gap behaves in a similar way as the order
parameter (\fig{fig:orderpar}), namely, without electron-lattice coupling it
changes smoothly across the phase transition, whereas the changes for finite
coupling are significantly sharper.

In \fig{fig:chargesusc} at $V=1.0$, a gapless excitation at
${\mathbf q}={\bf Q}$ can be seen, which occurs due to a
small but finite distortion. As already discussed in \sect{sec:statics}, this
distortion is finite only due to the finite-size effects and should be zero
in the thermodynamic limit. 

We note that the gap $\Delta_{C}$ in the charge spectrum is 
different from the one commonly used in DMRG calculations,
$\Delta :=[E_0(N+2)-E_0(N)]/2$, where $E_0(N+2)$ and $E_0(N)$ are the ground-state energies for 
systems consisting of $N+2$ and $N$ particles, respectively.
Indeed, as a function of $V$, $\Delta$ shows a behavior {\em opposite} to
$\Delta_{C}$, with $\Delta=0$ in the unbroken phase and $\Delta>0$ at large $V$.\cite{Vojta01}

\begin{figure}[t]
  \centering
  \includegraphics[width=0.45\textwidth]{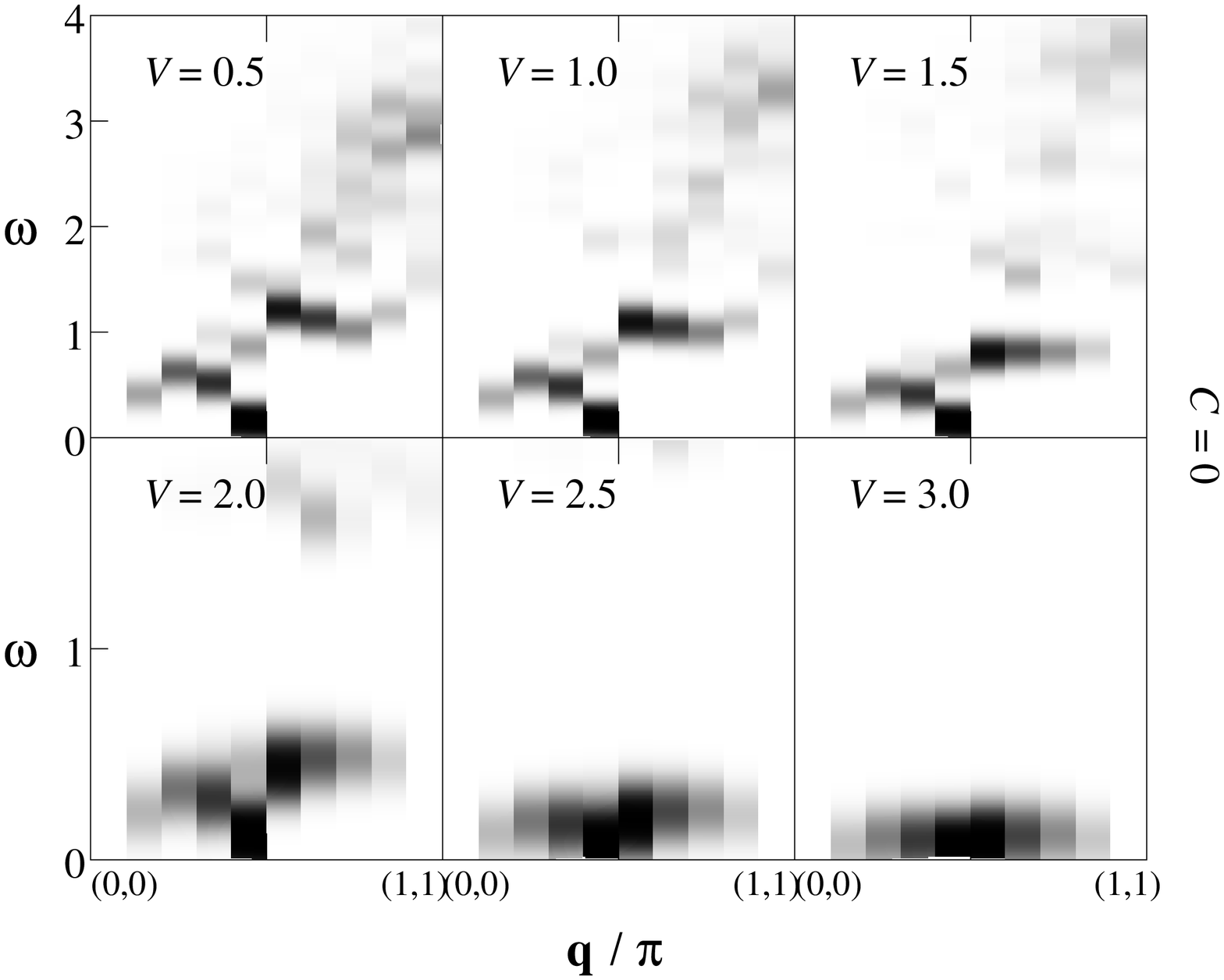}\\
  \includegraphics[width=0.45\textwidth]{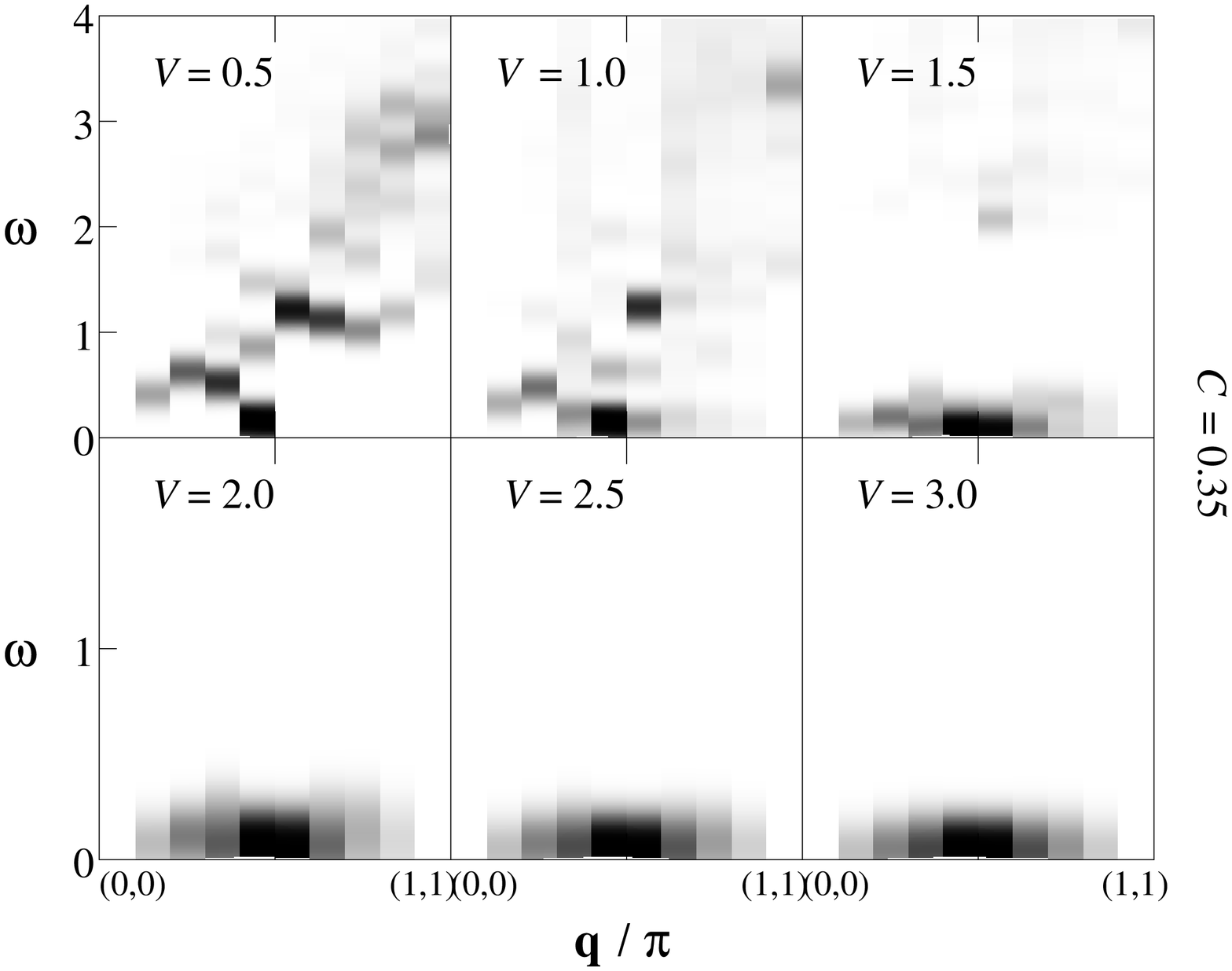}
  \caption{\label{fig:spinsusc}%
    Spin susceptibility calculated on an $8\times 2$ ladder. Upper panel: without lattice
    coupling. Lower panel: With lattice coupling ($C=0.35$). 
    Presentation as in \fig{fig:chargesusc}.}
\end{figure}
\begin{figure}[t]
  \centering
  \includegraphics[width=0.45\textwidth]{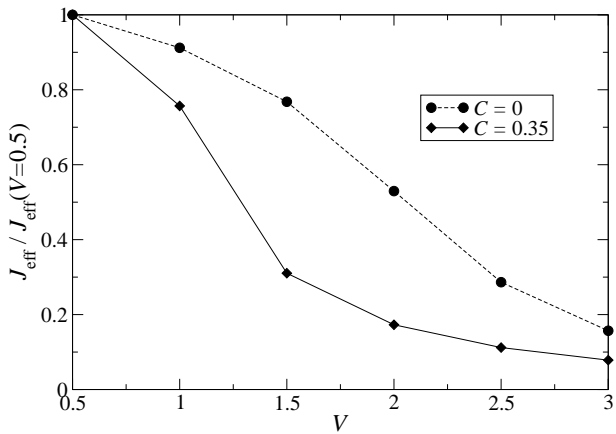}
  \caption{\label{fig:Jeff}%
    Effective magnetic exchange interaction $J_{\rm eff}$ in ladder direction
    in units of $J_{\rm eff}(V=0.5)$ as a
    function of $V$, extracted from the spin 
    susceptibility \eq{eq:spinsusz}. The interaction is shown with
    (solid line) and without lattice coupling (dashed line).
  }
\end{figure}

The spin susceptibility [\eq{eq:spinsusz}] calculated on the same system 
is shown in \fig{fig:spinsusc}. The momentum scan consists of two ranges,
from $(q_a,q_b)=(0,0)$ to $(0,\pi)$, and from $(\pi,0)$ to ($\pi,\pi$). 
In the first range ($q_a=0$)
one can clearly see the dispersion of an effective one-dimensional Heisenberg
model, as predicted by perturbation theory.\cite{Horsch98}
The main change of the spin susceptibility as a function of $V$ in this range 
is a decrease of the effective magnetic exchange interaction $J_{\rm eff}$ with increasing 
charge order. It can be extracted from the spin dispersion using
$J_{\rm eff}=2\,\omega(0,\pi/2)/\pi$ (Ref.~\onlinecite{Cloizeaux62}), and is
shown in \fig{fig:Jeff}. 
According to magnetic susceptibility
measurements,\cite{Weiden97} $J_{\rm eff}$ in the low-temperature phase 
is approximately 0.8 of the exchange in the disordered phase. 
Our results are in a good qualitative agreement with these 
data in the sense that an increase in charge order goes 
together with a decrease in $J_{\rm eff}$.\cite{Gros99}
However, a quantitative comparison   
cannot be made since in the experiment the ordering and, correspondingly, $J_{\rm eff}$,
are traced as a function of temperature
for given other system parameters while we investigate the ordering at $T=0$ as a function
of the extended Hubbard repulsion $V$. Our calculation also agrees well with the 
analytical results of Refs.~\onlinecite{Vojta01} and 
\onlinecite{Hubsch99}, where it was shown that the exchange
rapidly decreases with increasing $V$. Quantitatively, at $V=3,C=0$ our results give 
$J_{\rm eff}\approx 0.06$ while perturbation  theory  \cite{Vojta01,Hubsch99} 
predicts $J_{\rm eff}\approx 0.04$. Moreover, for $V=1.3,C=0.35$, which give 
a lattice distortion close to the value observed experimentally 
(see Fig. 4), the exchange parameter in Fig.13 is about 67\,meV 
which is very close to the experimental 60\,meV observed in the 
inelastic neutron scattering measurements of Ref.~\onlinecite{Grenier01}.

\begin{figure}[t]
  \centering
  \includegraphics[width=0.45\textwidth]{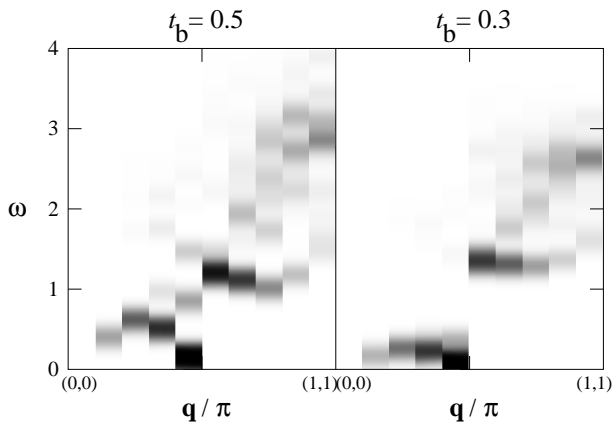}
  \caption{\label{fig:comp_tpar}%
    Spin susceptibility in the disordered phase at $V=0.5$ for hopping along
    the ladder $t_b=0.5$ (left) and $t_b=0.3$ (right). Momentum scan as in
    \fig{fig:chargesusc}.} 
\end{figure}

The second range in the spin spectrum (\fig{fig:spinsusc}),
with $q_a=\pi$, shows only high energy excitations at small $V$,
but again an effective Heisenberg dispersion at large $V$.
For small $V$ the gap in the spin spectrum
is very close to the charge gap, indicating that it is
due to charge excitations. To verify this conjecture, we calculated charge and
spin susceptibilities in the noninteracting limit $V=0$, with $t_b=0$ (isolated
rungs). In this case charge and spin susceptibilities are equal for
$q_a=\pi$ and the gap is exactly the difference between the bonding and the
antibonding state given by $2t_a$.  
Secondly, we analyzed the dependence of the spin
susceptibility on the hopping $t_b$ along the ladder in the disordered phase
at $V=0.5$ (\fig{fig:comp_tpar}). Whereas the dispersion
for $q_a=0$ scales as $t_b^2$, which is clear evidence
of the magnetic origin of these excitations, the difference between the
maximal and minimal excitation energy for $q_a=\pi$
scales as $t_b$. These observations show a direct interplay
between the spin and the dipole-active charge excitations, which is similar to the  
"charged" magnons introduced in Ref.~\onlinecite{Damascelli} for interpretation
of the infrared absorption spectra of NaV$_2$O$_5$. 

\begin{figure}[t]
  \centering
  \includegraphics[width=0.45\textwidth]{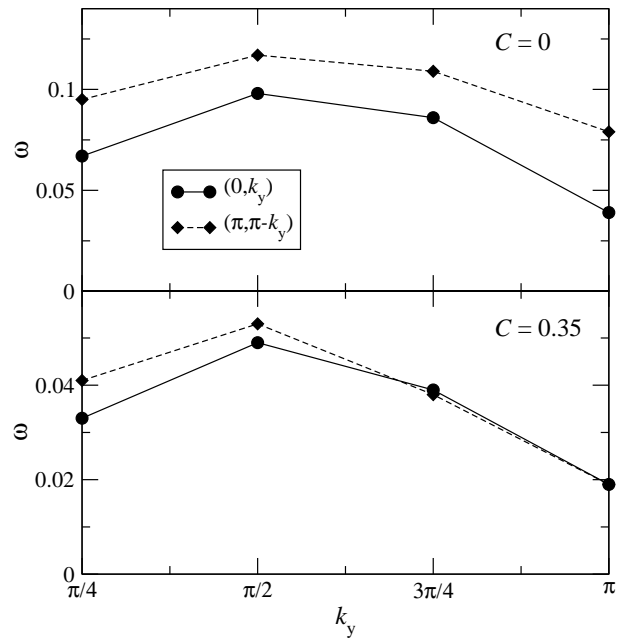}
  \caption{\label{fig:spindisp_v3}%
    Spin dispersion in the ordered phase at $V=3.0$ extracted from
    \fig{fig:spinsusc}. Bottom: with lattice coupling, top: without
    lattice coupling. Direction of momentum scans: Solid lines: $(0,0)\to(0,\pi)$, dashed
    lines: $(\pi,\pi)\to(\pi,0)$.
  }
\end{figure}

It is interesting to note that the spin spectra in the ordered phase appear to
possess a mirror symmetry with respect to the central tick mark in \fig{fig:spinsusc}.
To quantify this observation, the dispersions of
the low-energy excitations at $V=3.0$ have been depicted in \fig{fig:spindisp_v3}
The dispersions for $q_a=0$ and $q_a=\pi$
are indeed very similar. Without lattice coupling
the dispersion with $q_a=\pi$ is shifted upwards compared to $q_a=0$ because
of the small but finite charge gap at $V=3.0$ (see \fig{fig:chgap}). 
With lattice coupling and at interactions where no charge gap occurs the agreement is even better.
This behavior can be understood in the following way. In the disordered phase
where each electron on a rung occupies a molecular orbital consisting of two
sites, momenta $q=(0,\pi)$ and
$q=(\pi,0)$ are not equivalent (same spin on the two sites of the rung,
versus opposite spin on two sites of neighboring rungs). In this phase pure spin
excitations with $q_a=\pi$ are not possible since they require different
spins on different sites within a rung. This could be achieved only by
exciting another electronic state within the rung, which has the energy
$2t_a$.
In the totally zig-zag ordered
state where the electrons are located on one site of the rung, these
momenta become equivalent. The same holds for momenta
$q=(0,0)$ and $q=(\pi,\pi)$.

The overall effect of charge ordering on the dynamical susceptibilities can best be
seen by comparing the plots for $V=2.5, C=0.0$ and $V=1.5, C=0.35$, where the
values of the order parameter are similar, see \fig{fig:orderpar}. The spin and
charge excitations shown in these plots are qualitatively the same and the susceptibilities
differ only slightly on a few points.
From these figures we conclude that the dynamical spin and charge
susceptibilities of the system mainly depend on the order parameter but not
on the way in which the order has been achieved.  

%%%%%%%%%%%%%%%%%%%%%%%%%%%%%%%%%%%%%%%%%%%%%%%%%%%%%%%%%%%%%%%%%%%%%%%%%
\section{Hubbard-Holstein model}\label{sec:holstein}
%%%%%%%%%%%%%%%%%%%%%%%%%%%%%%%%%%%%%%%%%%%%%%%%%%%%%%%%%%%%%%%%%%%%%%%%%

So far we have only considered static distortions of the lattice. Although
this is a good approximation if the dynamic fluctuations
around these equilibrium positions are small, quantum phonon effects can
play an important role, especially in the critical region. In this section,
we therefore consider the extended Hubbard-Holstein model (EHHM)
\begin{equation}\label{eq:hamphon}
  H = H_{\rm EHM}+\sum_i\left[\frac{1}{2M}\hat p_i^2+\frac{\kappa}{2}\hat
  z_i^2-C\hat z_in_i\right],
\end{equation}
with $H_{\rm EHM}$ defined in \eq{hehm} and $M$ being the mass of the
local oscillators. The operators $\hat z_i$ and $\hat 
p_i$ are the coordinate and momentum of the ion on lattice site $i$, and all
other quantities are defined in \eq{ham}. 
When expressed in phonon creation and annihilation operators, it reads (up to a constant)
\begin{equation}\label{eq:HolsteinHubbard}
  H = H_\text{EHM}
      +\omega_0 \sum_i b^\dagger_ib^{\phantom{\dagger}}_i
      -g\sum_i (b^\dagger_i+ b^{\phantom{\dagger}}_i) n_i.
\end{equation}
Here $b^\dag_i$ ($b_i$)
creates (annihilates) a phonon of frequency $\omega_0$ (with $\hbar=1$) at lattice
site $i$, and the phonons are locally coupled to the electron density
with coupling strength $g=C\sqrt{\omega_0/2\kappa}$.

\begin{figure}[t]
  \centering
  \includegraphics[width=0.45\textwidth]{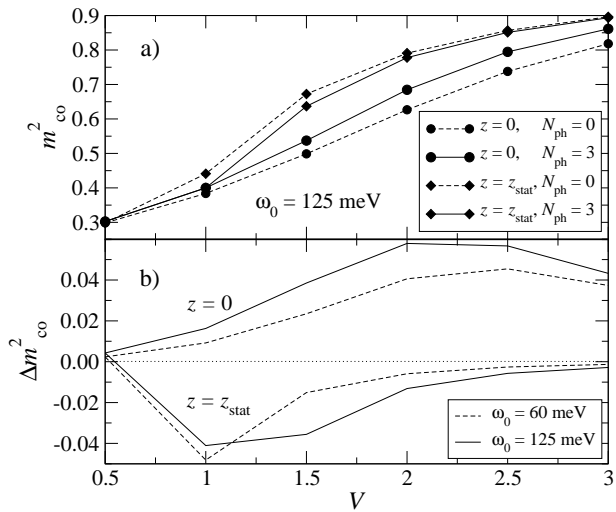}
  \caption{\label{fig:ehhm}%
    (a) Order parameter $m^2_{\rm CO}$, calculated on a cluster
    with four rungs. Phonon parameters are $C=0.35$, $\kappa=0.125$, and
    $\omega_0=125$\,meV. Parameters are chosen as shown in the caption. 
    (b) $\Delta m^2_{\rm CO}=m^2_{\rm CO}(N_{\rm 
    ph}=3)-m^2_{\rm CO}(N_{\rm ph}=0)$ for two phonon frequencies
    $\omega_0=60$ and $125$\,meV. Upper two curves: Without static
    distortion. Lower two curves: with static distortion. The dotted line
    marks $\Delta m^2_{\rm CO}=0$.
   }
\end{figure}

We would like to point out that the nature of the local phonon
mode in \eq{eq:HolsteinHubbard} is not specified and could correspond to one
of several phonon modes in the vanadates. 
Clearly, the use of dispersionless Einstein phonons
neglects any coupling between lattice distortions of neighboring
sites. However, a coupling of the Holstein type 
is the strongest phonon mode in NaV$_2$O$_5$.\cite{Spitaler03}
It also represents the simplest
model for electron-phonon interactions, and has been successfully
used to describe the physics of other transition metal oxides such as
the manganites.\cite{Edwards02}

Compared to exact diagonalization of the model described 
by the Hamiltonian in Eq.~(\ref{ham}), an additional
difficulty arises in the case of the EHHM since
the number of phonons is not conserved. Consequently, even for a finite
number of lattice sites, the Hilbert space contains an infinite number of
states, and has to be truncated in some way in order to apply the Lanczos
method. For this reason the size of the systems which 
can be investigated is also considerably reduced.  
We restricted ourselves to a lattice with four rungs and chose a subset
of the phonon states as \cite{Marsiglio93,Wellein96}
\begin{equation}\label{eq:basis}
  |r\rangle_\text{ph} =
  \prod_{i=1}^N\frac{1}{\sqrt{\nu_i^{(r)}!}}
  \left(b^\dag_i\right)^{\nu_i^{(r)}}
  |0\rangle_\text{ph}\,,
\end{equation}
where $\nu^{(r)}_i$ denotes the number of phonons at lattice site $i$ and
$|0\rangle_\text{ph}$ is the phonon vacuum state. Alternatively, the basis states
could also be formulated in momentum space as in Ref.~\onlinecite{Marsiglio93}.

Now the truncation of the Hilbert space consists of
restricting the total number of phonons in the $|r\rangle_\text{ph}$ subset as
\begin{equation}\label{eq:truncation}
\sum_{i=1}^N \nu^{(r)}_i\leq N_\text{ph}
\end{equation}
leading to $(N_\text{ph}+N-1)!/[N_\text{ph}!(N-1)!]$ allowed phonon
configurations for $N$ sites. We would like to point out that the set of basis 
states in Eqs.~(\ref{eq:basis}) and ~(\ref{eq:truncation}) consists
of all possible phonon states with up to $N_{\rm ph}$ phonons excited. 
In particular, the Hilbert space includes all linear combinations of such states .

Usually, $N_\text{ph}$ is increased until convergence of
an observable of interest $O$ is achieved. The latter can be monitored
by calculating the relative error
$|O(N_\text{ph}+1)-O(N_\text{ph})|/|O(N_\text{ph})|$.% As we shall see below,

Due to the complexity of the EHHM and the value of parameters, it is not
possible to include enough  phonon states to obtain converged
results. Nevertheless, as in the case of the pure Holstein model
\cite{Marsiglio95} it is still possible to deduce the tendency of the results
as $N_\text{ph}$ is increased and thereby obtain information about the exact 
results (corresponding to $N_\text{ph}=\infty$).

To reduce the required number of dynamical phonons, it is expedient
to introduce static distortions $z_i$ as a coordinate transformation
$\hat z_i=z_i+\hat x_i$ 
so that quantum fluctuations $\hat x_i$ take place around the position $z_i$. 
Applying this transformation to \eq{eq:hamphon} yields
\begin{align}
  H &= H_{\rm EHM}
     + \sum_i\left[\frac{\kappa}{2} z_i^2 - C z_i n_i \right]\nonumber\\
    &+ \sum_i\left[\frac{1}{2M}\hat p_i^2 + \frac{\kappa}{2}\hat x_i^2
                   + (\kappa  z_i - C n_i) \hat x_i \right],\label{eq:hamphontransf}
\end{align}
which in second quantization results in an expression analogous to
\eq{eq:HolsteinHubbard}. Note that the first 
line in the above equation is the same as \eq{ham}.
For the static distortions $z_i$ we
again use the zig-zag order pattern (\ref{eq_5}) and determine the optimal
value of $z$ by minimizing the ground-state energy in the presence of
phonons yielding a static distortion $z_{\rm stat}$, which is related to
$z_{\rm min}$ introduced in \sect{sec:statics} by $z_{\rm min}=z_{\rm
  stat}(N_{\rm ph}=0)$.

Note that we perform this coordinate transformation
only because the number of phonons accessible in our calculations is very
small, and in this case it is better to start from a different equilibrium
position $z=z_{\rm stat}$ and not from $z=0$. If it were possible to use
$N_{\rm ph}=\infty$, this coordinate transformation would have no influence on
the physical results and the actual lattice distortions would be produced
by the dynamical phonons as a coherent state of oscillators
associated with the ions, independent of any initial coordinate
transformation. Of course the broken symmetry would only 
occur in the thermodynamic limit, while correlations of the phonon positions
exist already on finite lattices.

The effect of dynamical phonons on the charge order parameter is shown in
\fig{fig:ehhm}. We did calculations for several values of $V$ at phonon
frequencies $\omega_0=60$\,meV and $\omega_0=125$\,meV, the two most relevant
modes in NaV$_2$O$_5$.\cite{Spitaler03} The smaller frequency belongs to a
collective vibration which includes displacements of the
vanadium and oxygen ions, whereas the larger one corresponds to a vibration of
the apical oxygen along the $z$ axis.

From the upper panel of \fig{fig:ehhm}
one can easily see that for the nontransformed coordinates (circles) the
inclusion of dynamical phonons with $\omega_0=125$\,meV considerably increases
the charge ordering. 
Calculations with $N_{\rm ph}=1$ and $2$ (not shown) revealed that the
increase is monotonic in the number of phonon states and we conclude that for
convergence many more phonon states would be necessary.
For the distorted lattice (diamonds), the dynamical phonons actually decrease
the charge ordering for $V\ge1.0$, and the strongest effect occurs in the
vicinity of the phase transition at $V=1.0$ and $V=1.5$. The reason for this
decrease is that $z_{\rm stat}$ is shifted downwards with
increasing number of dynamical phonons. At $V=1.0$, where a finite $z_{\rm
  stat}$ at $N_{\rm ph}=0$ is a finite size effect (\sect{sec:order}), $z_{\rm
stat}$ is reduced to zero for $N_{\rm ph}=3$. At $V\ge 1.5$ the relative
change in $z_{\rm stat}$ is similar to that in $m_{\rm CO}^2$ (\fig{fig:ehhm}).
We want to mention that the two solid curves in the upper panel of
\fig{fig:ehhm} give an upper and a lower boundary for the actual value of the
order parameter on the four rung lattice, since for $N_{\rm ph}\to=\infty$ 
results for $z=0$ and $z=z_{\rm stat}$ become equivalent, as discussed above.

In the lower panel of \fig{fig:ehhm} the difference $\Delta m^2_{\rm CO}$ of
the order parameter in 
the presence of dynamical phonons to $N_{\rm ph}=0$ is shown including data
for $\omega_0=60$\,meV. For $z=0$
(upper two curves) this deviation is positive and the 
effect is always larger for $\omega_0=125$\,meV, which corresponds to a
larger value of $g$. For $z=z_{\rm stat}$ 
(lower two curves) it is negative for $V \geq 1.0$. 
The crossing at $V=1.0$ is due to the 
small finite value of $z_{\rm stat}$ for $N_{\rm ph}=0$, i.e., a finite size effect.

\begin{figure}[t]
  \centering
  \includegraphics[width=0.45\textwidth]{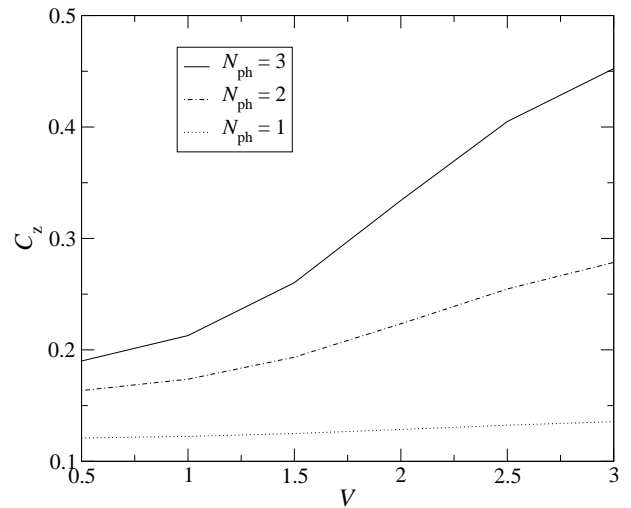}
  \caption{\label{fig:zcorr}%
    Correlation function $C_{\rm z}$ as defined in \eq{eq:zcorr} without static
    distortions for $\omega_0=125$\,meV. The number of phonons is given in
    the caption.} 
\end{figure}

It is interesting to study the order pattern of the dynamically induced
distortions. For this purpose we define the correlation function 
\begin{equation}\label{eq:zcorr}
  C_{\rm z}=\frac{1}{N^2}\sum_{ij}e^{i{\mathbf Q}({\mathbf R}_i-{\mathbf
  R}_j)}\Big\langle \big ( \hat z_i-\langle \hat z_i\rangle\big )
  \big (\hat z_j-\langle \hat z_j\rangle\big )\Big\rangle,
\end{equation}
which measures the zig-zag ordering of the lattice distortions, similar to
\eq{msqr} for the charge densities. For the nontransformed coordinates it is
depicted in \fig{fig:zcorr} for $\omega_0=125$\,meV and different numbers
of dynamical phonons. From this figure it is clear that
dynamical phonons induce zig-zag lattice distortions which strongly increase
around the phase transition point. Note that the correlation function $C_{\rm
  z}$ is not normalized to the interval $[0,1]$ since lattice distortions are
not conserved quantities, different from, e.g., charges. 

For the transformed coordinates we can calculate the dynamically induced
zig-zag distortions $z_{\rm dyn}$ directly from the expectation values
$\langle \hat z_i\rangle$ since the symmetry of the system is broken
explicitly. Results show that the sum $z_{\rm tot}=z_{\rm stat}+z_{\rm dyn}$ of the
static distortion and the 
dynamically induced distortion is always smaller than the value $z_{\rm min}$
determined in \sect{sec:order}, and this effect is most
pronounced near the phase transition. For $V=1.5$ and $N_{\rm ph}=3$ we got
$z_{\rm stat}=0.889$ and $z_{\rm dyn}=0.013$ yielding a total zig-zag distortion of
$z_{\rm tot}=0.902$, which is noticeably smaller than 
$z_{\rm min}=1.001$ for $N_{\rm ph}=0$. Well above the transition point the
dynamically induced distortions are very small, for instance for $V=3.0$ and
$N_{\rm ph}=3$ calculations gave $z_{\rm dyn}=0.0005$, and $z_{\rm
stat}=1.290$ is only slightly smaller than $z_{\rm min}=1.294$.

From this analysis we conclude that
using just static distortions gives qualitatively correct results but
overestimates the lattice distortions, in particular, in the vicinity of the
phase transition. The value of the order parameter in the full dynamical
model with $N_{\rm ph}=\infty$ 
will therefore be somewhat smaller than in \sect{sec:order}, as discussed above.

%%%%%%%%%%%%%%%%%%%%%%%%%%%%%%%%%%%%%%%%%%%%%%%%%%%%%%%%%%%%
\section{Conclusions}\label{sec:conclusions}
%%%%%%%%%%%%%%%%%%%%%%%%%%%%%%%%%%%%%%%%%%%%%%%%%%%%%%%%%%%%

In this paper we investigated the influence of lattice effects on the charge
ordering transition, the spectrum of kink excitations 
and dynamical susceptibilities of quarter-filled ladder systems 
and considered the $\alpha^\prime$-NaV$_2$O$_5$ compound as an example. 
For this purpose we modified the Hamiltonian of
the extended Hubbard model by terms which take into account the coupling
of electrons to lattice distortions.  
The lattice rigidity $\kappa$ and the Holstein coupling constant $C$ 
used in our model were
determined by first-principles band-structure calculations.
The physical properties were calculated 
by the exact diagonalization technique.
The results for the ground-state energy and the order
parameter show that by including static distortions, 
the phase transition is shifted significantly downward to lower values of the 
nearest-neighbor Coulomb
interaction $V.$ 
The calculated displacements of the vanadium ions 
due to the electron-lattice coupling
in the charge ordered phase are
in good agreement with experimental measurements.\cite{Luedecke99} 
We also found a virial theorem to be fulfilled to high precision for 
the terms in the Hamiltonian that couple to the lattice.

As low-energy excitations of the distorted ground state we considered kink
excitations, where the charge order pattern changes along the
ladder between two degenerate configurations. 
These kinks are long when $m_{\rm CO}^2$ is small, and become shorter with increasing order.
The kink lengths and energies at small $m_{\rm CO}^2$ 
are comparable with those of a classical $\phi^4$ model.\cite{Sherman01}

Moreover, we studied the extended Hubbard-Holstein model to
investigate the effect of dynamical phonons. Results showed that they have a
strong influence on the charge order parameter in the vicinity of the phase
transition. An analysis of the correlations of the dynamically induced
distortions revealed that phonons indeed favor zig-zag lattice
distortions. We showed that using just static distortions somewhat
overestimates the actual value of the lattice distortion, but well away from
the transition point this dynamical effect is very small and a description by static 
distortions gives already accurate results.   

In addition to these static properties  we
also calculated the dynamic charge and spin susceptibilities. We showed that
the main features of these quantities are determined by the value of the order parameter
and not by the way this value is achieved. From the spin susceptibility we
extracted the effective magnetic exchange interaction along the ladder,
which exhibits a pronounced decrease with increasing charge order. The magnitude 
of this parameter taken at $V=1.3,C=0.35$ is in good agreement with the experimental
inelastic neutron scattering data.\cite{Grenier01}

%%%%%%%%%%%%%%%%%%%%%%%%%%%%%%%%%%%%%%%%%%%%%%%%%%%%%%%%%%%%
\section*{Acknowledgments}
%%%%%%%%%%%%%%%%%%%%%%%%%%%%%%%%%%%%%%%%%%%%%%%%%%%%%%%%%%%%

This work has been supported by the Austrian Science Fund (FWF), project No. P15520.
M.A. and M.H. have been supported by the doctoral scholarship program of the
Austrian Academy of Sciences. 
One of us (E.Y.S) is grateful to P. Lemmens and M.N. Popova for interesting discussions.

%%%%%%%%%%%%%%%%%%%%%%%%%%%%%%%%%%%%%%%%%%%%%%%%%%%%%%%%%%%%

\end{document}